\documentclass{svjour2}
\smartqed
\usepackage{graphicx}
\journalname{JLTP}

\begin{document}

\title{Advanced concepts in Josephson junction reflection amplifiers}
\author{Pasi L\"ahteenm\"aki, Visa Vesterinen, Juha Hassel, G. S. Paraoanu, Heikki Sepp\"a and Pertti Hakonen}

\institute{
P. L\"ahteenm\"aki, G. S. Paraoanu, P. J. Hakonen \at
O.V. Lounasmaa Laboratory/LTL, P.O. Box 15100, FI-00076 AALTO, Finland \\
\and
V. Vesterinen, J. Hassel, H. Sepp\"a \at
VTT Technical Research Centre of Finland, PO Box 1000, FI-02044 VTT, Finland
}
\date{Received: date / Accepted: date}

\maketitle

\begin{abstract}
 Low-noise amplification at
microwave frequencies has become increasingly important for the research related to
superconducting qubits and nanoelectromechanical systems.
The fundamental limit of added noise by a phase-preserving
amplifier is the standard quantum limit, often expressed as noise
temperature $T_q$=$\hbar\omega/2k_B$. Towards the goal of the quantum
limit, we have developed an amplifier based
on intrinsic negative resistance of a selectively
damped Josephson junction. Here we present measurement results on previously proposed wide-band microwave amplification and
discuss the challenges for improvements on the existing designs.
We have also studied flux-pumped metamaterial-based parametric amplifiers,
whose operating frequency can be widely tuned by external DC-flux, and  demonstrate operation at $2\omega$ pumping, in contrast to the typical metamaterial amplifiers pumped via signal lines at $\omega$.
\end{abstract}

\section{Introduction}
The importance of quantum limited amplification has increased in
recent years due to growth of research related to superconducting
qubits and nanomechanical systems \cite{PROC2}. Several different approaches for making low-noise parametric
amplifiers based on Josephson junctions have been proposed and investigated \cite{PROC1}. The lowest noise
temperatures so far have been achieved using nondegenerate parametric amplifiers \cite{PROC3,PROC4,PROC5}. Other
implementations of near-quantum limited amplification have been realized by means of
Josephson ring oscillators \cite{PROC6}, DC-SQUIDs \cite{PROC7,PROC8}, and junction arrays \cite{PROC3,PROC9,PROC10,PROC11,PROC12}. Negative differential resistance devices similar to
our approach, but utilizing tunnel diodes have also been studied in the past \cite{PROC13}.
Our negative resistance device \cite{PROC1}, however, is unique in the respect that its high-gain operation
involves self-organisation. Self-organization improves the noise performance of the device, and much better
performance is achieved compared with other SQUID-type negative resistance amplifiers \cite{SQUID}. The basic scheme behind the wideband environment for the SJA is illustrated in Fig. 1 together with the schematics for a reflection-mode parametric amplifier.

\begin{figure}[tb]
\includegraphics[width=11cm]{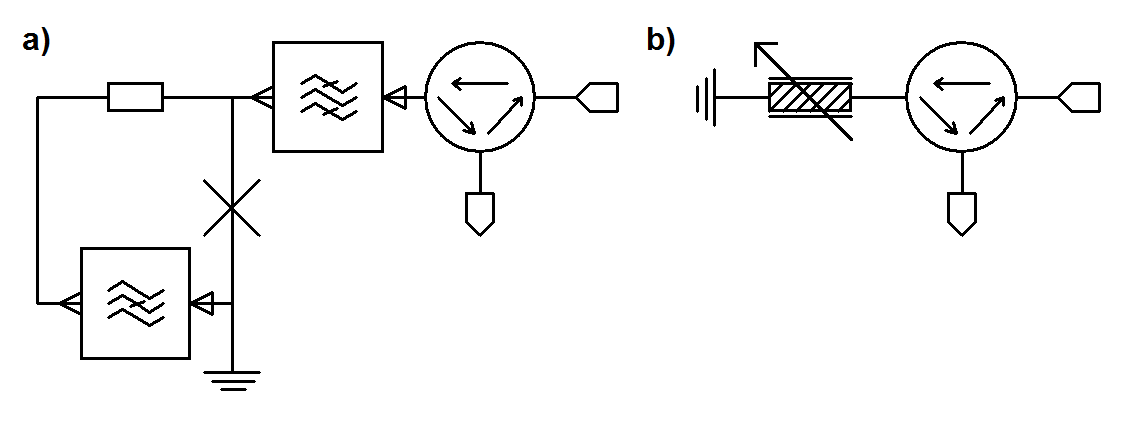}
\caption{a) Illustration of the principle for wideband operation of the single junction reflection amplifier (SJA). The on-chip environment, consisting here of a bandstop filter and a resistor, will shunt the off-signal-band frequencies while the signal frequencies are shunted by the external environment itself. During the reflection, the external signal experiences gain due to the effective negative AC-resistance of the DC-biased junction (biasing circuits not drawn). The band-splitting windows can be wideband as long as damping over all frequencies is sufficient to stabilize the high-frequency dynamics responsible for the negative resistance of the system.
b) Basic configuration of a reflection-mode parametric amplifier. The amplifier is formed by a resonator whose resonant frequency is strongly modulated, either via the signal line by an RF-pump at $\omega$ or through an external coil at $2\omega$ (common for single SQUID parametric amplifiers). Our metamaterial devices can be operated both at $\omega$ via the signal line and at $2\omega$ using on-chip coils.}
\end{figure}

The second topic of this paper, our metamaterial based parametric amplifier is flexible and tunable over wide range of
frequencies due to the fact that it consists of 250 SQUIDs embedded in a coplanar waveguide resonator whose resonant frequency can be tuned by external flux. A similar sample was used in the experimental observation of the dynamical Casimir effect \cite{PNAS}. This modulation of flux affects the critical current of the SQUIDs and through this effect changes the Josephson inductance. The combined nonlinearity of these SQUIDs, relative to the resonant frequency of the cavity, is much higher than modulation of the resonant frequency that can reliably be accomplished in any simpler design with smaller
number of junctions or SQUIDs.\footnote{This is due to the fact that, within the range of available junction parameters in our fabrication process, the tunability of resonant frequency of a single SQUID resonator would be strongly limited  by the necessary, non-ideal tuning capacitance.}

An external flux can be used to modulate the resonant frequency of the cavity at twice the
signal frequency, which results in parametric gain. Josephson junction metamaterial amplifiers have been successfully demonstrated in Ref. \cite{PROC9}, but they have not been operated using $2\omega$ pumping before. This kind of pumping allows simpler measurement arrangements due to the absence of a strong pump tone close to the measurement band.


\section{Principles}

Negative resistance provides a simple concept for making high-gain reflection
amplifiers \cite{BOOK}. Reflection amplifiers in a microwave system with $Z_0=50$ $\Omega$
yield a power gain
$|S_{11}(\omega)|^2=|\Gamma(\omega)|^2$ which is determined by the
reflection coefficient
$\Gamma(\omega)=(Z_{in}(\omega)-Z_0)/(Z_{in}(\omega)+Z_0)$. Consequently,
when the impedance of the active component $Z_{in}$ equals $-Z_0$ the
nominal gain diverges. $Z_{in}$  can be engineered nearly frequency independent over large
bandwidths, which means that wide-band amplifiers can be realized using
components with negative resistance.

\begin{figure}[tb]
\includegraphics[width=5.5cm]{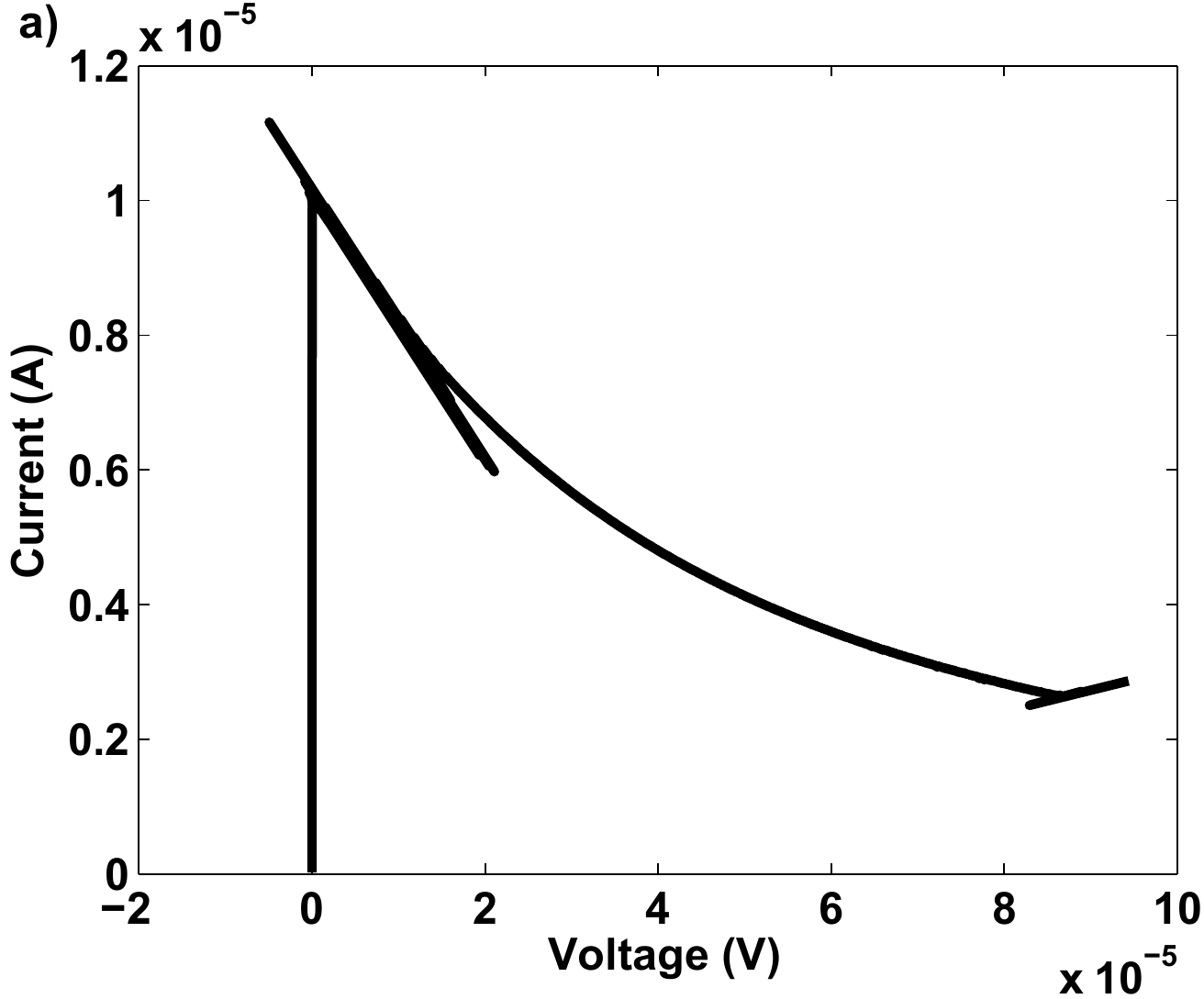}
\includegraphics[width=5.5cm]{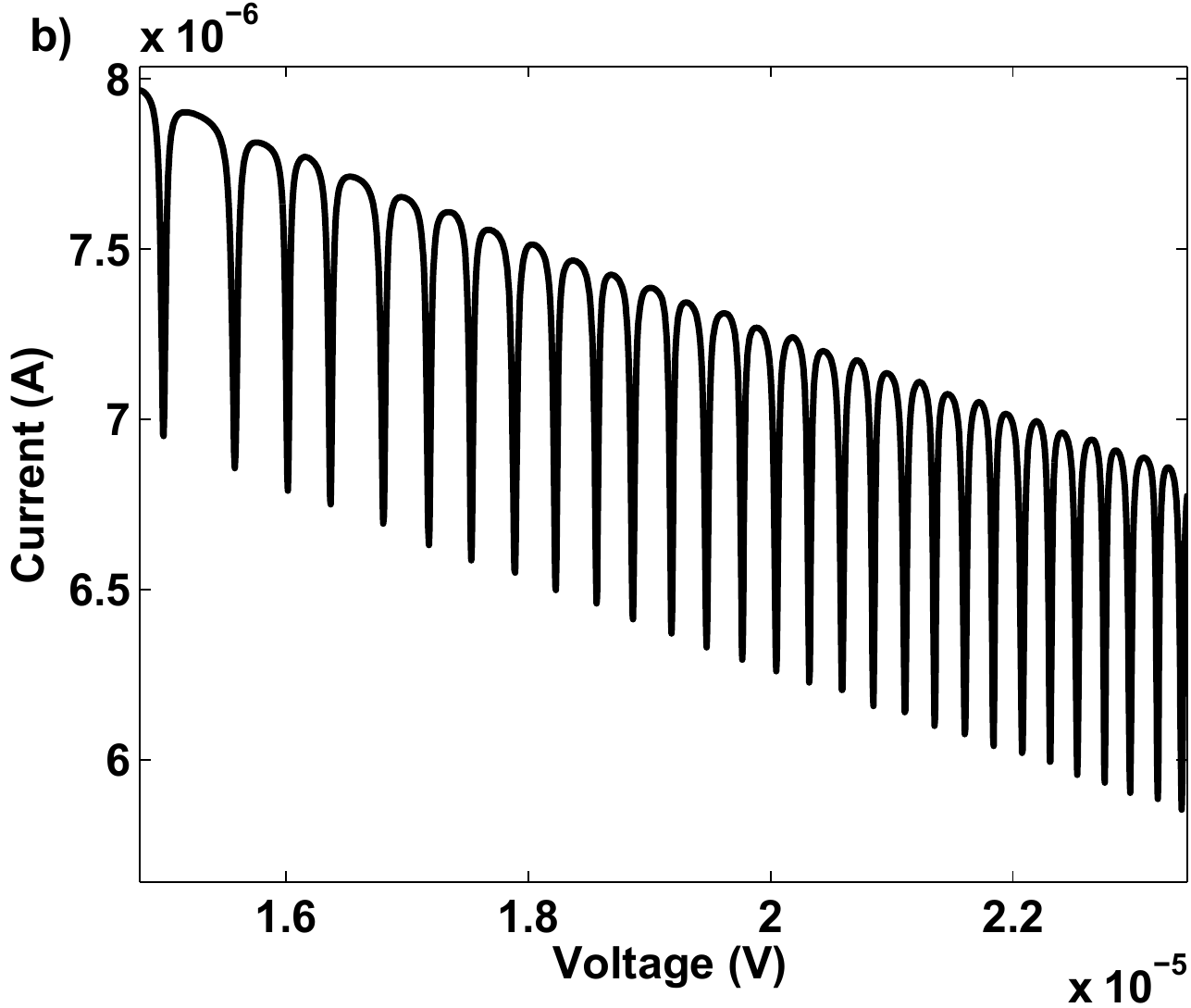}
\caption{a) Simulated IV-curve for $I_C$ = 10 $\mu$A, C = 0.1 pF junction
biased in 5 $\Omega$ environment (low impedance due to simpler simulation and illustration purposes). The discontinuous parts (two nearly straight lines)
are due to finite simulation time and slow Josephson dynamics in those regions of
the IV-curve. b) Part of the same IV-curve as in a), but with finite sweep time
and limited filtering. The high frequency dynamics, which causes the
negative resistance slope of the IV-curve is now clearly visible.}
\end{figure}
Josephson junctions can by employed to produce dynamic negative resistance
$R_d < 0$ at frequencies well below the Josephson frequency. The  dynamics of
the phase across the Josephson junction can be described by the resistively
and capacitively shunted junction model \cite{Tinkham}. Averaging over the
Josephson oscillations, an effective IV curve emerges, which is illustrated in Fig. 2. 
using a time-domain circuit simulation. As the voltage is ramped up linearly on a resistively shunted junction,
the average current \textit{through} the  Josephson junction decreases.

In order to stabilize the $R_d < 0$ of the junction for amplification,
a frequency selective shunt is introduced. This can be done,  e.g.  using
an LCR-bandstop filter. Because the direct noise from the shunt is
excluded in this sort of design, the only contribution to the noise of this
device originates from the mixed down noise from the Josephson
frequency \cite{PROC14}.

Since the negative resistance has no fundamental bandwidth
limitations except the limits due to the Josephson frequency itself and the
superconducting gap, it is principally possible to achieve
a very high gain-bandwidth product (GBW) as long as the negative
resistance branch of the junction can be stabilized by the signal
environment.

For the metamaterial amplifier, the velocity of signals in the coplanar waveguide depends on the applied flux: this dependence comes via the Josephson inductance of the SQUIDs which has flux dependence owing to the tunable critical current of the SQUID loops. The SQUIDs are densely
packed within the wavelength of the signals propagating in this medium, so that we are able to consider the
metamaterial in the continuum limit. The amplifier is operated by modulating the resonance frequency of this cavity at twice the
signal frequency, which results in parametric gain. The cavity is formed by  coupling the metamaterial
transmission line  weakly into the external world through a small on-chip capacitor, while the other end of the transmission line
is shorted.

\section{Measurements and the samples}

Our  single junction reflection amplifiers (SJA) were Nb-Al/AlOx-Nb junction devices manufactured using in-situ deposited tri-layer films  \cite{seppa}.
This  technique has been very successful in fabricating large area high quality tunnel junctions, but it is demanding to apply such a process for high quality sub micron tunnel junctions, which would be optimal for the low-noise single junction amplifiers \cite{PROC1}.
 Owing to the single junction structure,  shielding against stray magnetic fields does not need to be as efficient as for regular SQUIDs. The design parameters were the following: critical current $I_C=24.5$ $\mu$A,  shunt resistance  $R_s \approx 4$  $\Omega$, junction capacitance $C_J=390$ fF, and $C=4.3$ pF, $L=650$ pH for the band stop filter. The measured $I_C$ was $17$ $\mu$A.

\begin{figure}[tb]
\includegraphics[width=5.5cm]{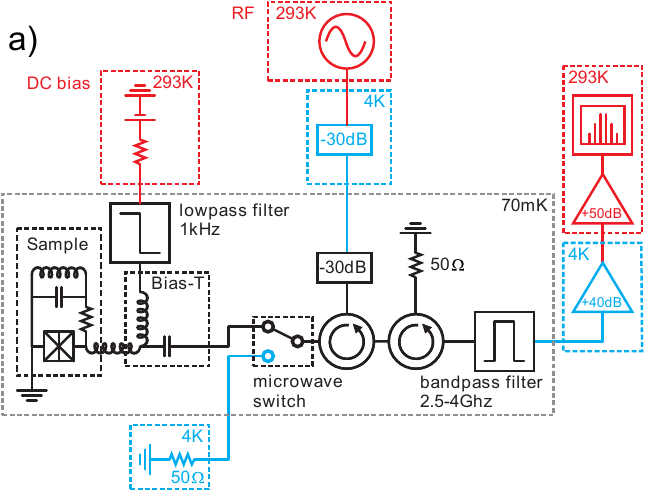}
\includegraphics[width=5.5cm]{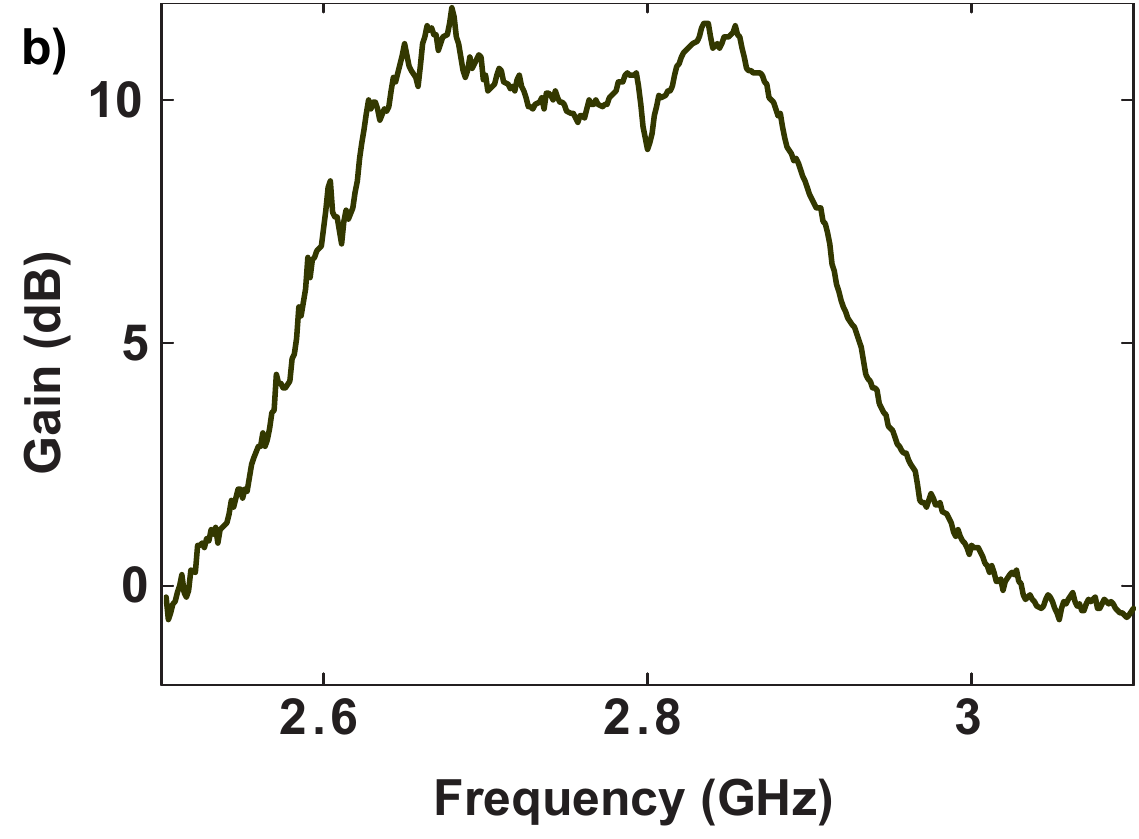}
\caption{a) The schematic for the sample and the components of
the measurement setup located at different temperatures. b) Measured $S_{11}$ (reflection gain) of the device
demonstrating wide-band gain. }
\end{figure}
The  schematics for SJA measurements  is illustrated in Fig. 3a. 
A bias-T was employed to provide DC-bias to the junction through a heavily filtered line. A microwave switch was employed for connecting a 50 $\Omega$ thermal noise source for calibration purposes. The cold low-noise HEMT amplifier had a noise temperature of ~$\sim 5$ K. Two circulators in series were employed to prevent the noise from the preamplifier from interfering with the investigated junction.

The measurement results demonstrating gain over a wide band are presented in
Fig. 3b. The linear gain is on the order of 10 
dB over frequencies between 2.6 GHz and 2.8 GHz.
The device
\begin{figure}[b]
\begin{center}
\includegraphics[width=10cm]{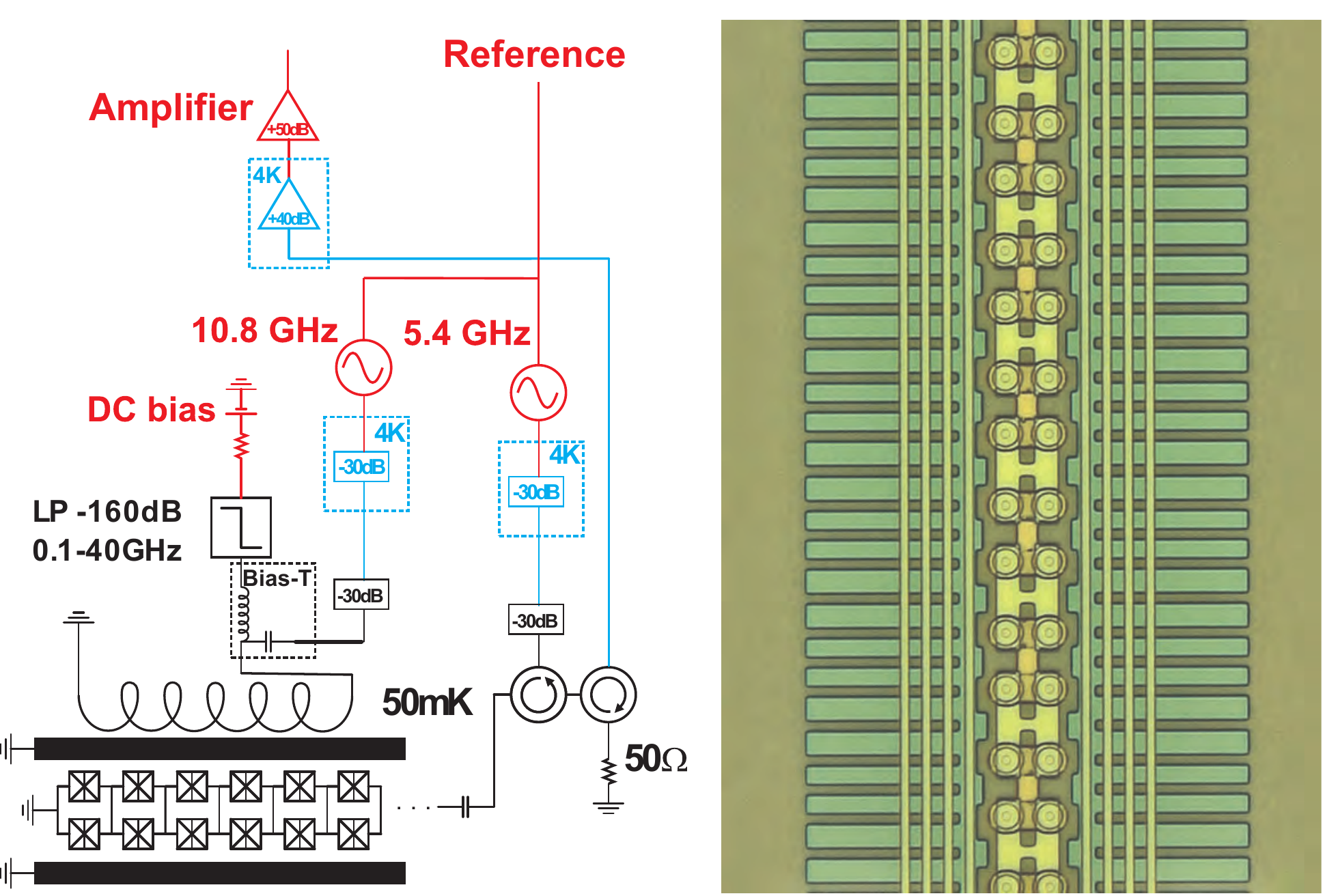}
\caption{Left: The modulated metamaterial sample and the measurement schematics. The pump is at twice the signal frequency. Right: Microscope image of the sample, in which 14 SQUID loops are visible in the central (vertical) conductor line. The pumping coil is visible as four vertical lines on both sides of the SQUID array. }
\end{center}
\end{figure}
measured in Ref. [2] had a voltage GBW = 40 MHz, while here we obtain
 GBW = 500 MHz. This was accomplished by operating the device
at lower bias voltage resulting in lower $\omega_J$ which, in turn,
results in a higher GBW \cite{PROC1}. However, as a result of the lower gain of the
amplifier in the present experiments, the noise compression mechanism \cite{PROC1}, deduced as previously from SNR-measurements, is significantly weakened and the noise added by the amplifier is around 10-20 quanta.

Our metamaterial amplifier was fabricated using a process similar to the single
junction amplifiers, but with a slightly lower critical current of $\sim 10$
$\mu$A. The employed measurement setup and an optical image of the  amplifier are depicted in Fig. 4. The measured reflection gain ($S_{11}$) is 
plotted in Fig. 5a at a few values of pump power. 

We observed in the degenerate parametric amplifier mode a maximum gain of around 24 dB in the in-phase
quadrature and a deamplification of $\sim 8$ dB in the out-of-phase quadrature (see Fig. 5b). The 2D 
quadrature plot when amplifying only noise without a probe signal is presented in Fig. 5c; 
it displays clearly the character of the phase sensitive gain on top of the
measurement system noise of 6-7 K.
\begin{figure}[tb]
\includegraphics[width=6cm]{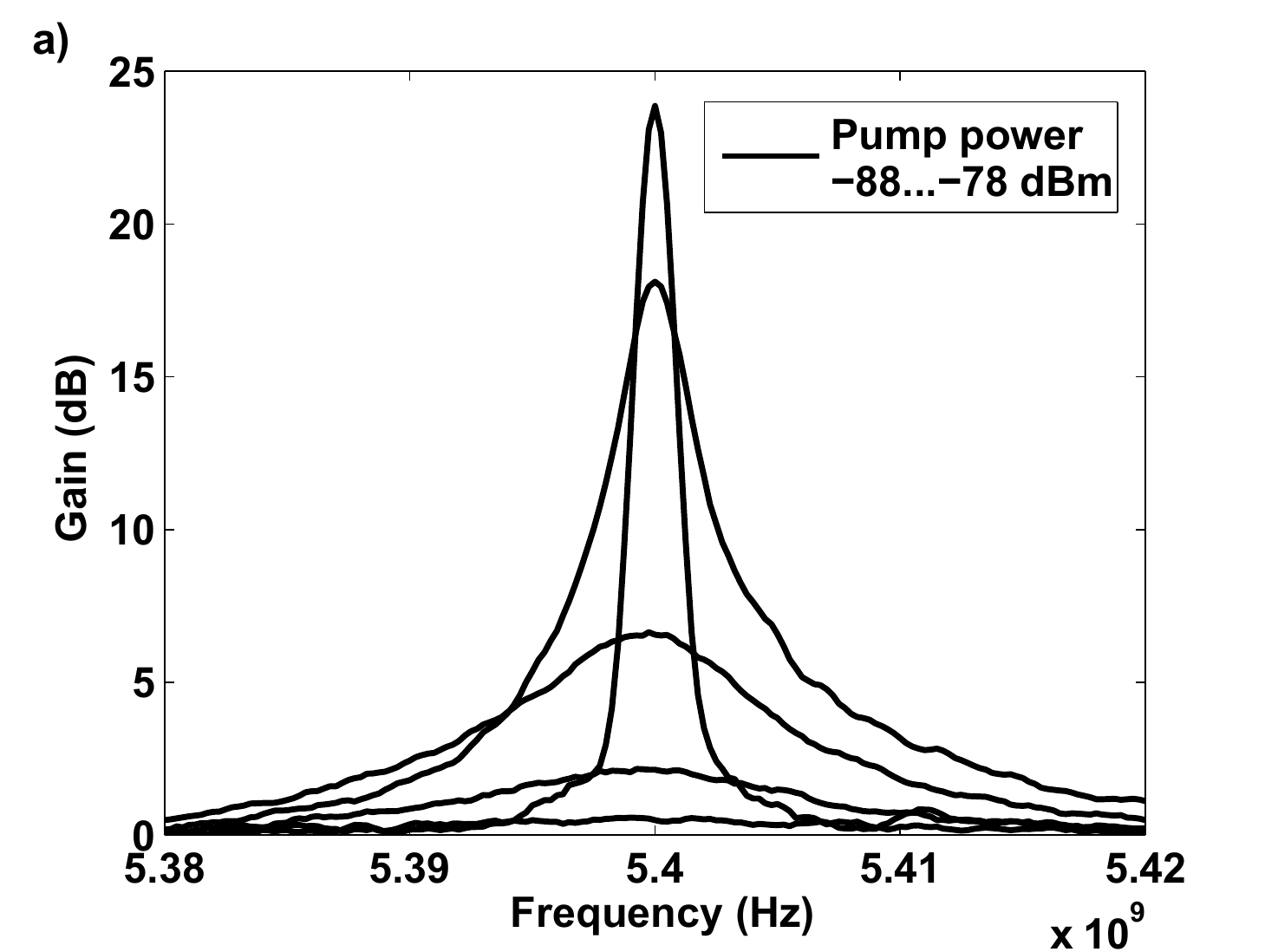}
\includegraphics[width=6cm]{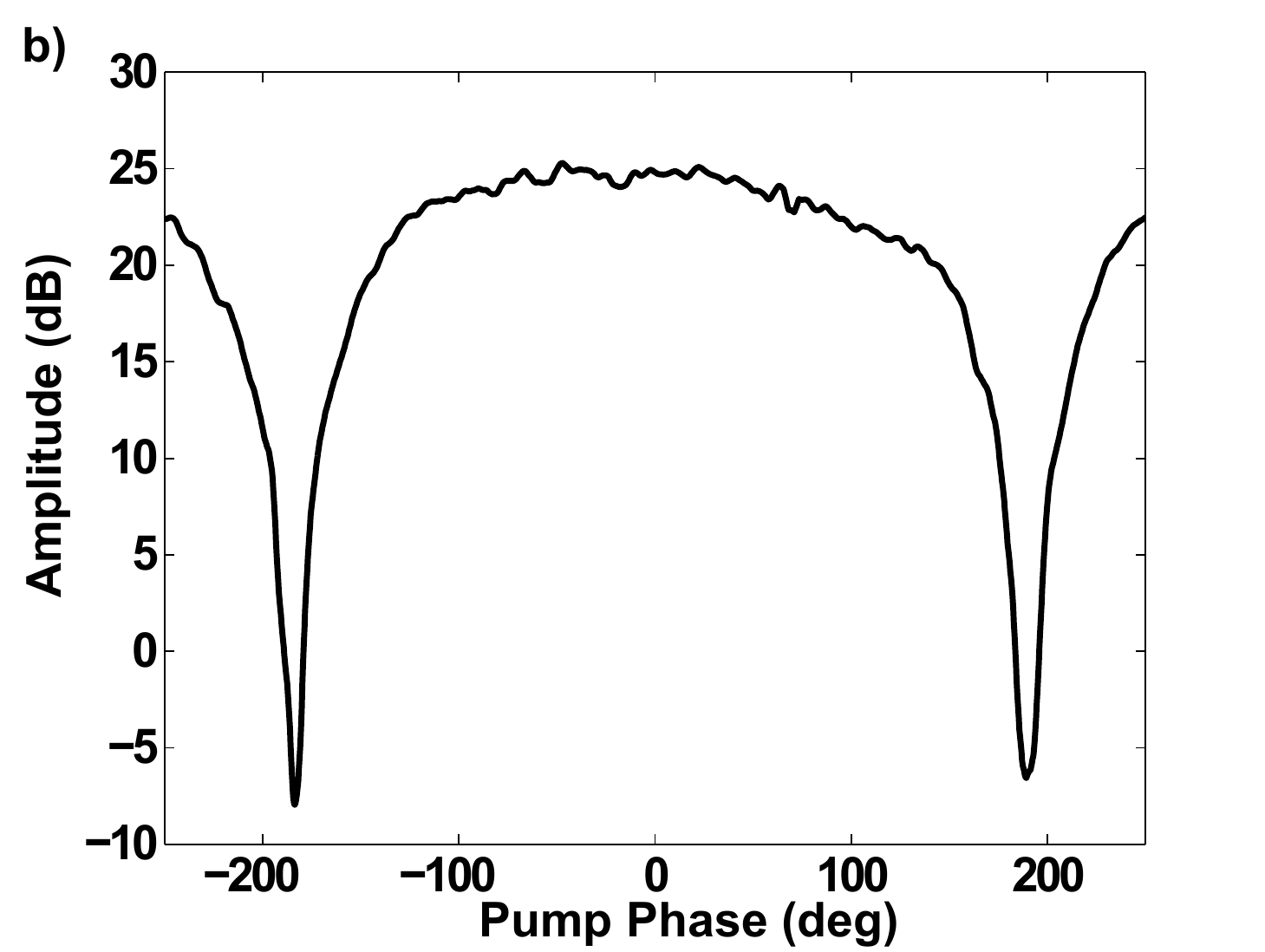}
\includegraphics[width=6cm]{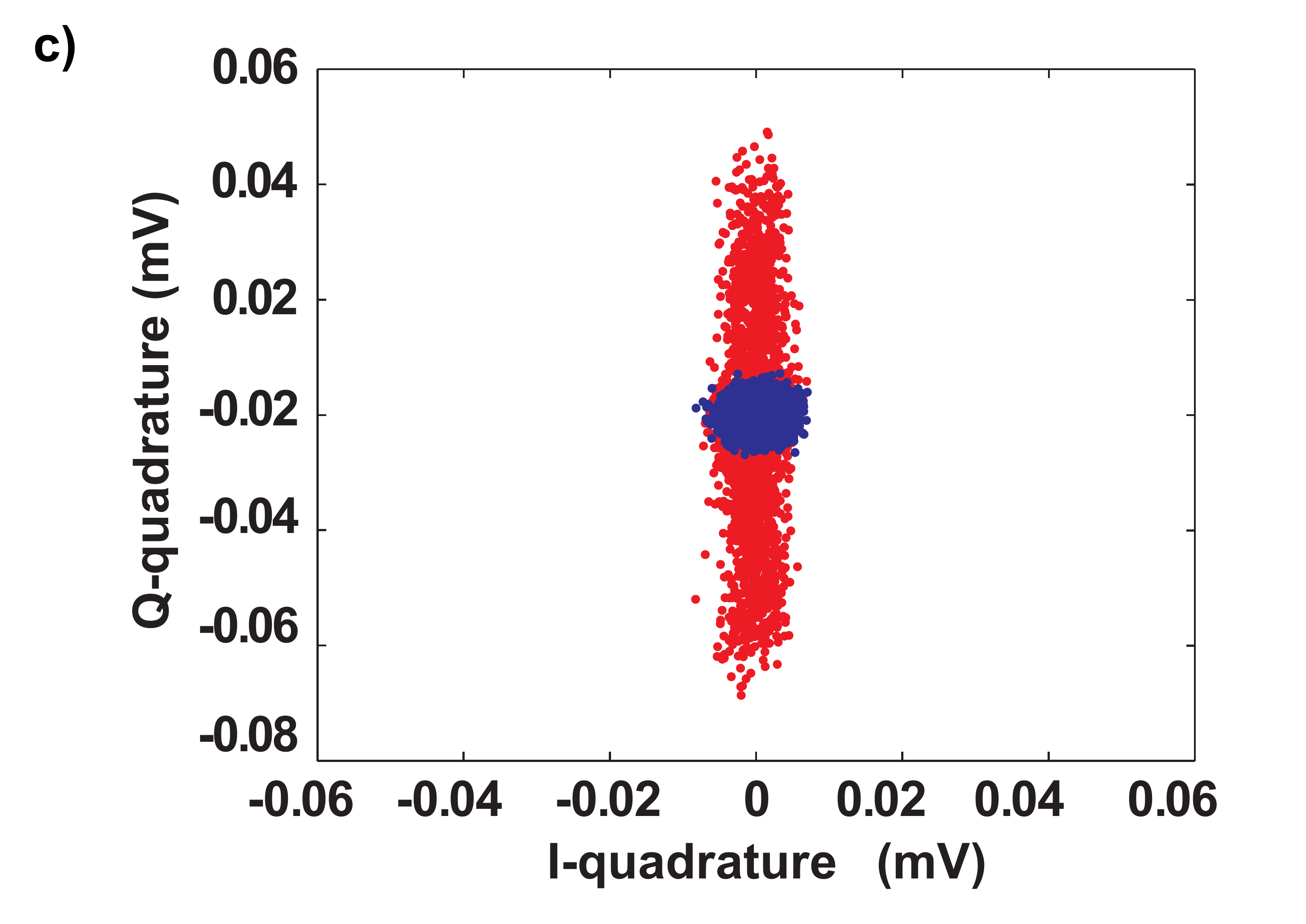}
\includegraphics[width=6cm]{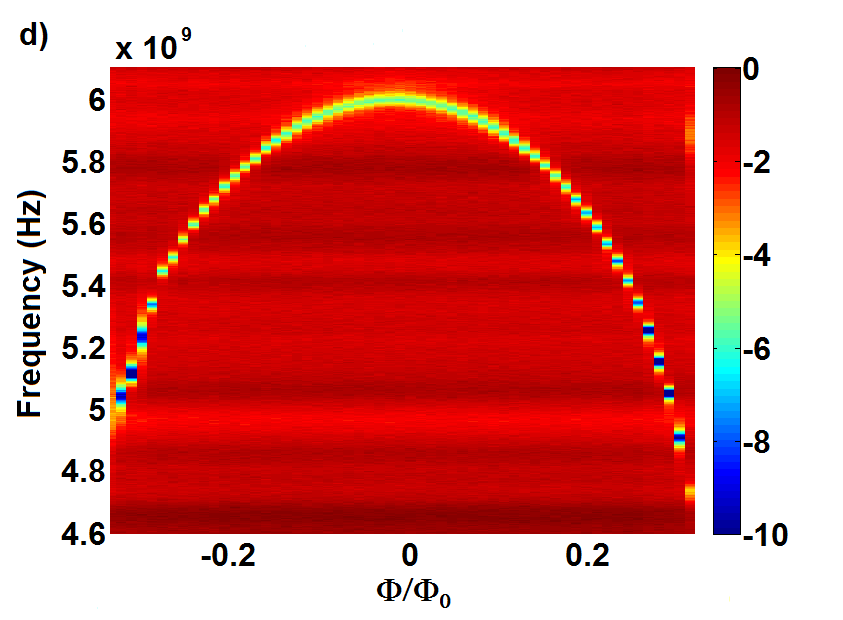}
\caption{a) In-phase gain at five different  levels of pump power around $-88\ldots-78$ dBm. b) Gain while the pump phase is rotated. c) Quadrature amplitude plot (blue - pump off, red - pump on).  d) Reflection magnitude vs. magnetic flux (in units of flux quanta $\phi_0$) shows how the resonance frequency can be tuned. The color scale on the right is given in dB units.}
\end{figure}
The observed resonance frequency vs. magnetic
flux in Fig. 5d indicates that the device can be tuned by approximately  1 GHz between 5 and 6 GHz. 
By tuning below 5 GHz, the device could no longer be operated properly, most likely due to increased microwave
losses or trapping of flux. The added noise energy deduced from the improvement
in the signal/noise ratio was $(1.5 - 2.0) \hbar \omega$ at best.


\section{Discussion}

The gain of our SJA devices originates from the mixing of the Josephson
oscillations ($\omega_J$) with the signal $\omega_S$
($\omega_S<<\omega_J$). The DC-bias-generated Josephson oscillations act as pump frequency in a parametric amplifier, which produces mixing from $\omega_S$ to side bands (idler frequencies) $\omega_J \pm \omega_S$ and back. With a large enough impedance at the down mixed frequency, voltage amplification is produced. In other words, the mixing results in an effective negative
resistance leading to gain of the signals. Consequently, wide-band operation via the mixing process requires proper control of the impedance environment over the signal band. The bandwidth of the Josephson oscillation is of no concern for the wideband operation as long as it remains well above the signal frequencies as is the case here at the employed bias points.

The Lorentzian form of gain in Ref. [2] results from the signal frequency LC-resonator
environment and the impedance matching circuit used in the device. However, such
resonator elements are not necessarily needed, and instead of the shunt circuit, one could
use any kind of filter. For example, a high-order rectangular
band stop filter could  be entirely shunted by
the source, in which case normal rules for limiting the band width of
the device do not apply. Even in the case of the LC-circuit, the
band width is a function of the bias voltage ($v_b$), and its magnitude can be changed
to some extent by tuning the bias. 

According to the RSJ model, the noise temperature of a Josephson junction parametric amplifier is  $(T_N)_{min} \approx \frac{\omega_s}{\omega_c}T$, ($\omega_s\ll\omega_c$), where the characteristic frequency $\omega_c = 2\frac{R_N}{L_J}$ is governed by the normal state resistance of the junction $R_N$ and the Josephson inductance $L_J$ \cite{likharev}. Using similar linearized analysis, a rather poor performance for negative-resistance junction amplifiers is predicted \cite{NEGAT}. According to our analytical calculations for selectively shunted devices \cite{PROC1},  the relevant reduction factor in the noise compression region becomes of form $\propto \frac{\omega_c}{\omega_J}$ with additional phase variation dependent factors. Altogether, the optimization of our SJA amplifiers is rather tricky and analytical estimates have to be verified by numerical simulations.

In the investigated SJA device here, we have in addition to the LC band stop
filter also an external LC-circuit as detailed in Fig. 6. The latter LC-circuit 
acts both as a matching circuit (slightly detuned) and as a shunting
circuit for the Josephson junction. With these LC-circuits having
rather small $Q$-factors ($Q\sim20$), wide-band operation could be realized. However,
the maximum observed power gain was nearly 100 times smaller than in measurements
with a single, high-$Q$ band resonator \cite{PROC1}. In order to enhance this gain for reaching the noise compression regime,
we have investigated devices with smaller critical currents and developed tunable matching circuits \cite{Visa}.

\begin{figure}[ht]
\centering
\includegraphics[width=5cm]{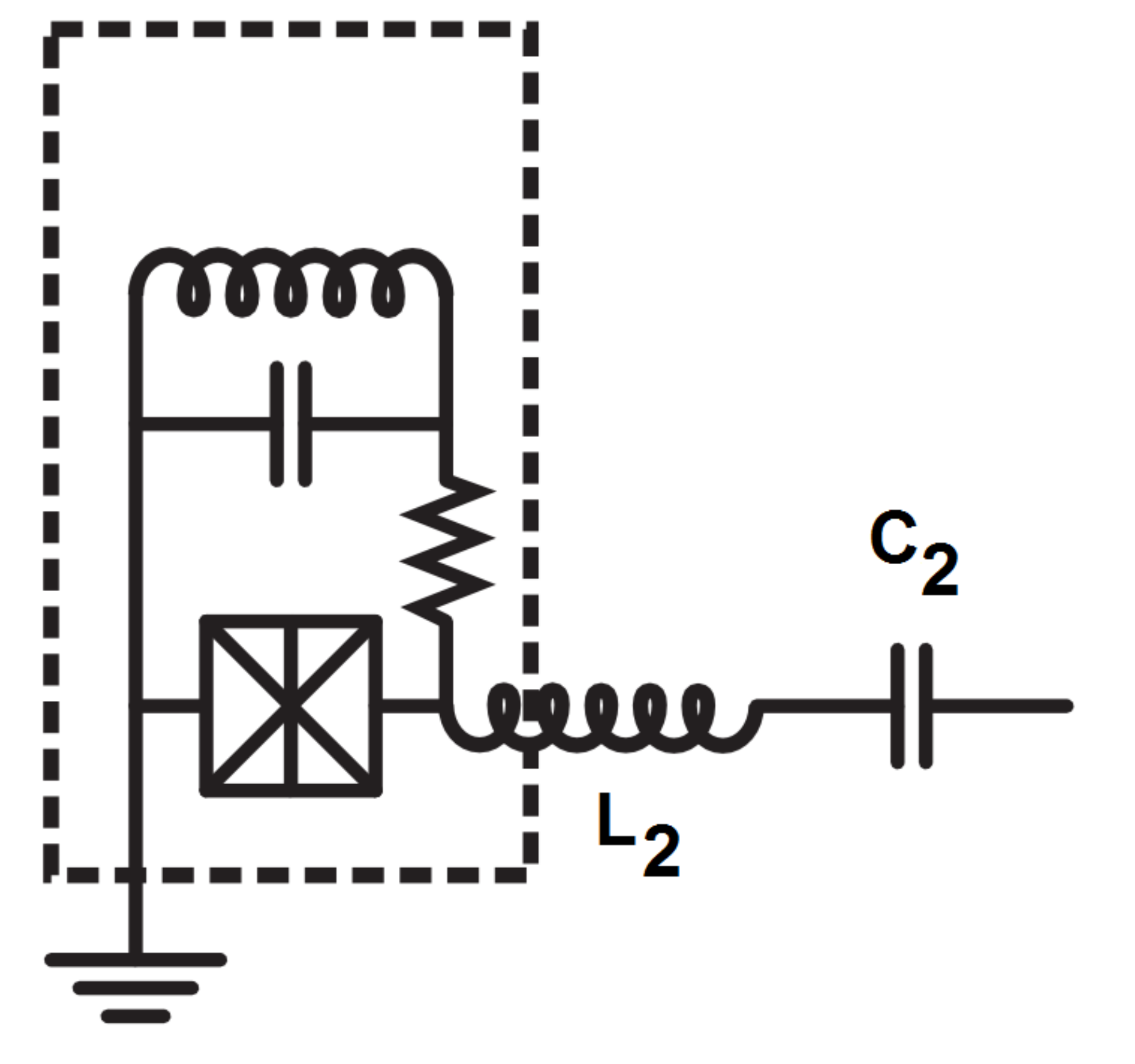}
\caption{Closeup of the schematic of the sample and the matching circuit.}
\end{figure}

As a final note on our single junction amplifier design, we want to emphasize that,
unlike typical non-degenerate parametric amplifiers, our design has two idler frequencies.
Without noise compression, both idlers will contribute by $\frac{1}{2} \hbar \omega$ to the noise.
What happens to these contributions with the noise compression is an open theoretical question at present, and it should
be resolved using methods beyond the semiclassical approximations employed in our work.

Our metamaterial amplifiers demonstrate that it is feasible to build a widely tunable
high gain amplifier which is operated by $2\omega$ pumping. However, the
existing samples still suffer from spurious resonances in the pump coil and from
microwave losses \cite{piers,SupSci} which should be addressed in the future
fabrication schemes. The microwave losses are not so troublesome at small power,
but they become severe at the pumping levels needed to obtain significant gain out of the metamaterial amplifiers.

\section{Conclusions}
We have presented experimental results which demonstrate wide-band gain of microwave amplifiers
based on the negative resistance of a single Josephson junction. Our
results are well backed by circuit simulations where the quantum noise is
taken into account as a semiclassical source term. As an integral part of the SJA operation, there is noise compression via self-organization,
 which requires careful engineering of the frequency-dependent environmental impedance seen by the junction.

We have also been first to demonstrate parametric gain in long, $2\omega$ flux-pumped Josephson metamaterial,
widely tunable over GHz frequencies. This
method of pumping is most convenient because it does not interfere with the signal frequencies, as may happen with
$\omega$ pumped amplifiers due to the strong signal at the center of the gain band. For our current samples, the $2\omega$ pumping does not work ideally and further development, especially in minimizing dielectric losses, is necessary in order to reach  quantum limited performance.

\begin{acknowledgements}
  Our work was supported in part by the EU 7th Framework Programme (FP7/2007-2013, Grant No. 228464 Microkelvin) and by the Academy of Finland (projects no. 250280 [LTQ CoE grant], no. 135908 and no. 263457).  This research project made use of the Aalto University Cryohall infrastructure.
\end{acknowledgements}


\begin{thebibliography}{100}

\bibitem[1]{PROC2}
A. A. Clerk, M. H. Devoret, S. M. Girvin, F. Marquardt, and R. J. Schoelkopf, \textit{Rev. Mod. Phys.} \textbf{81}, 1155 (2010).

\bibitem[2]{PROC1}
P. L\"ahteenm\"aki, V. Vesterinen, J. Hassel, H. Sepp\"a, and
P. Hakonen, \textit{Sci. Rep.} \textbf{2}, 276 (2012).

\bibitem[3]{PROC3}
M. Hatridge, R. Vijay, D. H. Slichter, J. Clarke, and I. Siddiqi,
\textit{Phys. Rev. B} \textbf{83}, 134501 (2011).

\bibitem[4]{PROC4}
B. Yurke, L. R. Corruccini, P. G. Kaminsky, and L.W. Rupp,  \textit{Phys. Rev. A} \textbf{39},
2519 (1989).

\bibitem[5]{PROC5}
B. Yurke, M. L. Roukes, R. Movshovich, and A. N. Pargellis, \textit{Appl. Phys. Lett.} \textbf{69}, 3078 (1996).

\bibitem[6]{PROC6}
N. Bergeal, F. Schackert, M. Metcalfe,	et al.  \textit{Nature} \textbf{465}, 64 (2010).

\bibitem[7]{PROC7}
M. M\"uck, J. B. Kycia, and J. Clarke, \textit{Appl. Phys. Lett.} \textbf{78}, 967 (2001).

\bibitem[8]{PROC8}
S. J. Asztalos, G. Carosi, C. Hagmann, et al. \textit{Phys. Rev. Lett.} \textbf{104}, 041301 (2010).

\bibitem[9]{PROC9}
M. A. Castellanos-Beltran, K. D. Irwin, G. C. Hilton, L. R. Vale, and K. W.
Lehnert,  \textit{Nature Phys.} \textbf{4}, 929 (2008).

\bibitem[10]{PROC10}
J. D. Teufel,  T. Donner, M. A. Castellanos-Beltran, J. W. Harlow, and K. W. Lehnert, \textit{Nature Nanotech.} \textbf{4},
820 (2009).

\bibitem[11]{PROC11}
M. A. Castellanos-Beltran and K. W. Lehnert, \textit{Appl.
Phys. Lett.} \textbf{91}, 083509 (2007).

\bibitem[12]{PROC12}
L. Spietz, K. Irwin, and J. Aumentado,  \textit{Appl. Phys.
Lett.} \textbf{93}, 082506 (2008).

\bibitem[13]{PROC13}
J. J. Tiemann, \textit{Proc. IRE.} \textbf{8}, 1418 (1960).

\bibitem[14]{SQUID}    A. Vystavkin, V. Gubankov, L. Kuzmin, K. Likharev, V. Migulin, V. Semenov, \emph{Magnetics, IEEE Trans.} \textbf{13},
233, (1977); and N. Calander, T. Claeson, S. Runder,  \emph{J.
Appl. Phys.} \textbf{53}, 5093 (1982).

\bibitem[15]{PNAS}
P. L\"ahteenm\"aki, G. S. Paraoanu, J. Hassel, and P. J. Hakonen, \textit{Proc. Natl. Acad. Sci}, \textbf{110}, 4234-4238 (2013).

\bibitem[16]{BOOK}
L. A. Blackwell and K. L. Kotzebue, \textit{Semiconductor Diode Parametric Amplifiers} (Prentice-Hall, Englewood Cliffs, 1961).

\bibitem[17]{Tinkham} M. Tinkham, \textit{Introduction to Supercondictivity, 2nd ed.} (Dover Publications Inc, Mineola, 2004)

\bibitem[18]{PROC14}
R. H. Koch, D. J. Van Harlingen, J. Clarke, \textit{Phys. Rev. B}, \textbf{26}, 74 (1982).

\bibitem[19]{seppa} H. Sepp\"a, M. Kiviranta, A. Satrapinski, L. Gr\"onberg, J, Salmi, and I. Suni, \textit{IEEE Trans. Appl. Supercond} \textbf{3}, 1816 (1993).

\bibitem[20]{likharev}
K. K. Likharev, \emph{Dynamics of Josephson Junctions And Circuits, 3rd ed.} (Gordon And Breach Publishers, Amsterdam, 1997).

\bibitem[21]{NEGAT} A. Vystavkin, V. Gubankov, L. Kuzmin, K. Likharev, V. Migulin, V. Semenov, Magnetics, IEEE Trans. \textbf{13}, 233 (1977).

\bibitem[22]{Visa} V. Vesterinen, J. Hassel, and H. Seppä, IEEE Trans. Appl. Supercond. \textbf{23}, 1500104 (2013).

\bibitem[23]{piers} J. Hassel, P. Lahteenmäki, A. Timofeev, G. S. Paraoanu, and P. J. Hakonen,
\textit{PIERS Proceedings}, 484 - 488, August 12-15, Stockholm, 2013.

\bibitem[24]{SupSci} D. Gunnarsson, J-M. Pirkkalainen, J. Li, et al. \textit{Sup. Sci. Techn.} \textbf{26}, 085010  (2013).


\end{thebibliography}
\end{document}